\documentclass[fleqn,usenatbib]{mnras}

\usepackage{amsfonts}

\usepackage[T1]{fontenc}
\usepackage{ae,aecompl}

\usepackage{grffile}
\newcommand\HII{{H\,\textsc{ii}}} 
\usepackage{etoolbox}
\makeatletter 
  \patchcmd{\NAT@citex}
    {\@citea\NAT@hyper@{%
      \NAT@nmfmt{\NAT@nm}%
      \hyper@natlinkbreak{\NAT@aysep\NAT@spacechar}{\@citeb\@extra@b@citeb}%
      \NAT@date}}
    {\@citea\NAT@nmfmt{\NAT@nm}%
    \NAT@aysep\NAT@spacechar\NAT@hyper@{\NAT@date}}{}{}

  \patchcmd{\NAT@citex}
    {\@citea\NAT@hyper@{%
      \NAT@nmfmt{\NAT@nm}%
      \hyper@natlinkbreak{\NAT@spacechar\NAT@@open\if*#1*\else#1\NAT@spacechar\fi}%
        {\@citeb\@extra@b@citeb}%
      \NAT@date}}
    {\@citea\NAT@nmfmt{\NAT@nm}%
    \NAT@spacechar\NAT@@open\if*#1*\else#1\NAT@spacechar\fi\NAT@hyper@{\NAT@date}}
    {}{}
\makeatother
\usepackage{graphicx}	
\usepackage{amsmath}	
\usepackage{amssymb}	

\title[Cooling of primordial gas]{Effect of lithium hydride on the cooling of primordial gas}

\author[B. Liu and V. Bromm]{Boyuan Liu\thanks{E-mail: boyuan@utexas.edu} and Volker Bromm
\\
Department of Astronomy, University of Texas, Austin, TX 78712, USA\\
}

\date{Accepted XXX. Received YYY; in original form ZZZ}

\pubyear{2017}

\begin{document}
\label{firstpage}
\pagerange{\pageref{firstpage}--\pageref{lastpage}}
\maketitle

\begin{abstract}
We complete the formulation of the standard model of first star formation by exploring the possible impact of $\mathrm{LiH}$ cooling, which has been neglected in previous simulations of non-linear collapse. Specifically, we find that at redshift $z\gtrsim 5$, the cooling by $\mathrm{LiH}$ has no effect on the thermal evolution of shocked primordial gas, and  of collapsing primordial gas into minihaloes or relic \HII\ regions, even if the primordial lithium abundance were enhanced by one order of magnitude. Adding the most important lithium species to a minimum network of primordial chemistry, we demonstrate that insufficient $\mathrm{LiH}$ is produced in all cases considered, about $[\mathrm{LiH/Li}]\sim 10^{-9}$ for $T\lesssim 100\ \mathrm{K}$. Indeed, $\mathrm{LiH}$ cooling would only be marginally significant in shocked primordial gas for the highly unlikely case that the $\mathrm{LiH}$ abundance were increased by nine orders of magnitude, implying that {\it all} lithium would have to be converted into $\mathrm{LiH}$. In this study, photo-destruction processes are not considered, and the collisional disassociation rate of $\mathrm{LiH}$ is possibly underestimated, rendering our results an extreme upper limit. Therefore, the cooling by $\mathrm{LiH}$ can safely be neglected for the thermal evolution of Population~III star-forming gas.
\end{abstract}



 
\section{Introduction}
\label{s1}
Star formation in general crucially depends on the cooling processes active in the host cloud \citep[e.g.][]{mckee2007}. Together with the dynamical evolution, the cooling physics determines the path through temperature-density ($T-n$) phase space of a self-gravitating gas cloud prior to fragmentation, and thus the mass scale of the pre-stellar core, approximately given by the Jeans mass, $M_{\mathrm{J}}\propto T^{3/2}n^{-1/2}$ (e.g. \citealt{palla2002}). For the first generation of stars, the so-called Population~III (Pop III), formed in minihaloes of total mass $\sim 10^{6}\ M_{\odot}$ at redshifts $z\gtrsim 20$, the current `standard model' has identified molecular hydrogen as primary cooling agent \citep[reviewed in][]{bromm2013}. Specifically, in the absence of metal species and dust, the primordial gas evolves adiabatically during the initial collapse into the minihalo potential, and is subsequently cooled by $\mathrm{H_{2}}$, built up to an abundance of $x_{\mathrm{H_{2}}}\sim 10^{-3}$, to $\sim 200\ \mathrm{K}$ at $n\sim 10^{4}\ \mathrm{cm^{-3}}$
(e.g. \citealt{bromm1999,bromm2002,abel2002,clarke2003}). The standard picture has to be augmented when considering the thermal and chemical behaviour of primordial gas in shocks encountered during the merging of dark matter haloes in early structure formation, in supernova (SN) blast waves \citep[e.g.][]{mackey2003,machida2005}, or in the presence of a cosmic-ray (CR) background, produced by early SN explosions \citep{cr2007,cr2016}.
In either case, the primordial gas retains a higher non-equilibrium fraction of free electrons, which in turn catalyzes an enhanced abundance of $\mathrm{H_{2}}$. The primordial gas can then cool to $\lesssim 100\ \mathrm{K}$ (e.g. \citealt{1987ApJ,kang1992}).
At that point, additional cooling is provided by hydrogen deuteride ($\mathrm{HD}$), which may enable the shock-compressed, primordial gas to cool to the temperature of the cosmic microwave background (CMB), $T_{\mathrm{CMB}}\gtrsim 30\ \mathrm{K}$ at $z\gtrsim 10$, within a Hubble time
\citep{larson1998,johnson2006}.   

Other than $\mathrm{H_{2}}$ and $\mathrm{HD}$, it has long been suspected that lithium hydride ($\mathrm{LiH}$) may also be an important coolant in primordial gas, owing to its large electric dipole moment \citep[e.g.][]{1984ApJ,stancil1996,bougleux1997,galli1998,bovino2011}. Previous studies have shown that the contribution of $\mathrm{LiH}$ to the total cooling rate is insignificant in the diffuse intergalactic medium (IGM) at $\gtrsim 10$, although it may play a role in collapse calculations \citep{lepp2002,galli2013}.
However, no detailed study has yet been conducted to assess the effect of $\mathrm{LiH}$ cooling for setting the initial conditions of fragmentation and star formation, during the non-linear stages of structure formation, when primordial gas collapses into newly virialized haloes.

In light of this, we specifically study the effect of $\mathrm{LiH}$ cooling in three situations: shocks from SN, collapse into minihaloes, and collapse inside relic \HII\ regions \citep[e.g.][]{greif2009}. In each case, our focus is on establishing an upper limit for the potential effect. In particular, we wish to find out whether including $\mathrm{LiH}$ can cool the gas (faster) to the CMB temperature prior to protostellar core formation.
In our model, the primordial lithium abundance is a key parameter, which is still subject to significant uncertainty, as it suffers from the well-known `cosmic lithium problem': the value measured in the atmospheres of metal-poor stars is three times lower than that predicted by standard Big Bang Nucleosynthesis (BBN) calculations (e.g. \citealt{asplund2008,spite2012}). A variety of explanations for this discrepancy have been put forward, generally from three perspectives: astrophysics, nuclear physics, or new physics beyond the standard BBN model (e.g., as reviewed by \citealt{iocco2009,fields2011}). In the third class of explanations, the primordial lithium abundance is perturbed from the standard BBN value by, e.g., decaying dark matter, changing fundamental constants, or large-scale inhomogeneities in the cosmic density. Note that such perturbations can also enhance the $\mathrm{Li}$ abundance. We therefore also investigate the case in which the primordial abundance of lithium is higher than the standard BBN value by one order of magnitude, in accordance with our overall goal of establishing a firm upper limit for the effect. The outline of this paper is as follows. In Sections~\ref{s2}, we describe our dynamical models. In Section~\ref{s3}, we present the results for maximal cooling (Section~\ref{s3.1}), and cooling from the full lithium chemical network (Section~\ref{s3.2}). Finally, we briefly summarize our findings in Section~\ref{s4}.

\section{Dynamical model}
\label{s2}
\subsection{Overview}
\label{s2.1}

\begin{figure*}
\centering
\includegraphics[width=2.0\columnwidth]{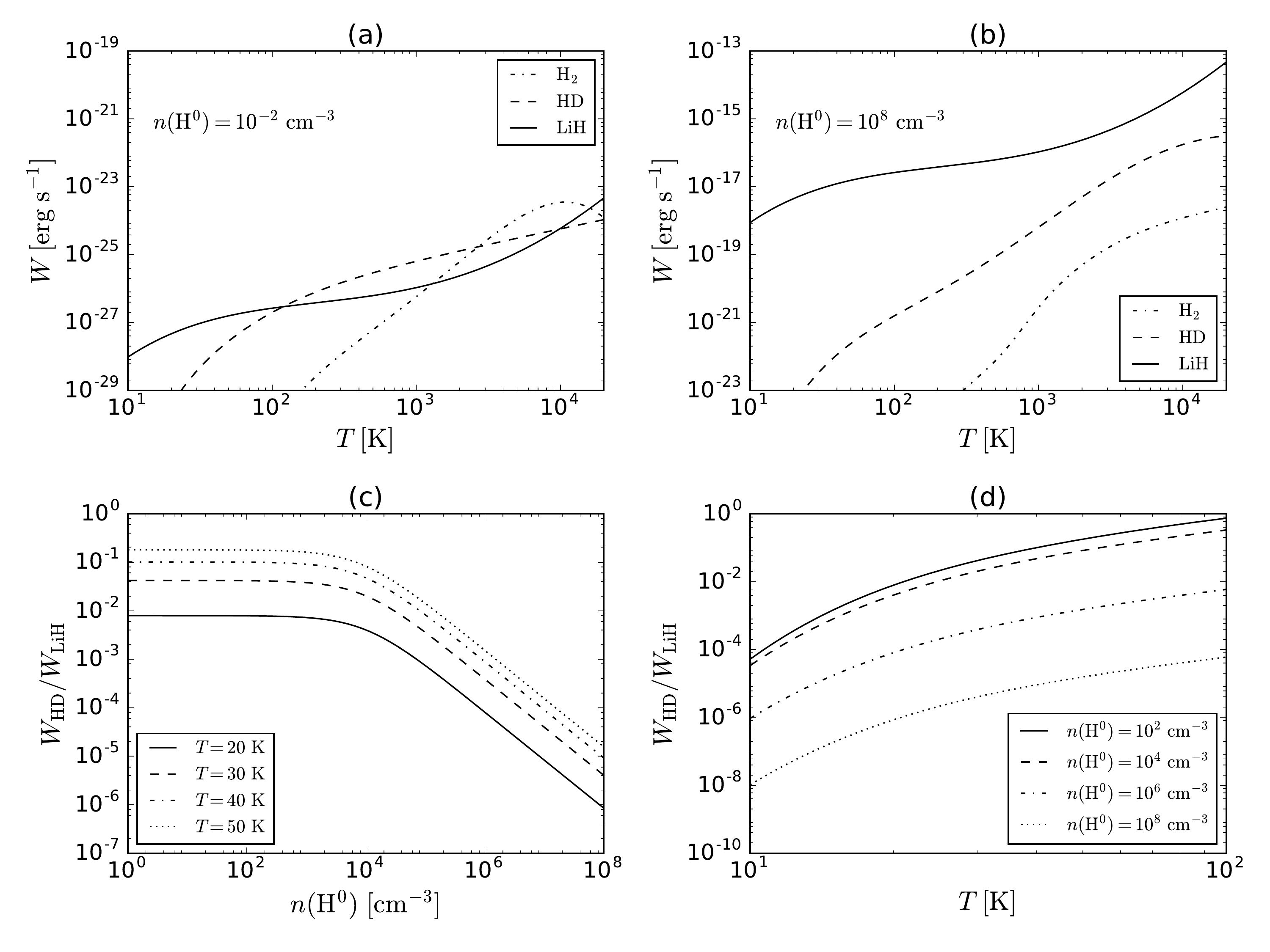}
\caption{Cooling functions in primordial gas. {\it Top panels}: cooling rates (per coolant particle) for the molecular coolants $\mathrm{H_{2}}$ (dashed-dotted), $\mathrm{HD}$ (dashed) and $\mathrm{LiH}$ (solid), in {\it (a)} the low-density regime with $n(\mathrm{H^{0}})=10^{-2}\ \mathrm{cm^{-3}}$, and {\it (b)} the high-density regime with $n(\mathrm{H^{0}})=10^{8}\ \mathrm{cm^{-3}}$. {\it Bottom panels}: ratio of the cooling rate (per coolant particle) for $\mathrm{HD}$ and $\mathrm{LiH}$, as a function of {\it (c)} collider number density, $n(\mathrm{H^{0}})$, and {\it (d)} gas temperature, $T$. In the former case, curves are plotted for different temperatures, and in the latter for different collider number densities.} 
\label{f1}
\end{figure*}

Similar to \citet[][JB06 henceforth]{johnson2006}, in order to explore the cooling process, we focus on the chemical and thermal evolution of primordial gas with simplified dynamics in different astrophysical contexts at high redshifts. To better assess the potential effect of $\mathrm{LiH}$ cooling, we specifically consider clouds at $z\sim 5$, so that the CMB temperature has sufficiently dropped to render  $\mathrm{LiH}$ competitive with $\mathrm{HD}$ cooling\footnote{It is not completely unrealistic to have gas clouds of primordial abundances at $z=5$, as the mixing of metals in the IGM is inhomogeneous and inefficient \citep[e.g.][]{tornatore2007,galli2013,jaacks2017}.}. This switch in the dominant cooling channel occurs for gas temperatures $T\lesssim 30\ \mathrm{K}$ (see Figure~\ref{f1}), a regime that is thermodynamically inaccessible at higher redshifts. We consider three environments in our calculations:
\begin{itemize}
\item (1) SN-shocked primordial gas, which is heated and ionized immediately after the shock, and evolves close to isobaric thereafter \citep{1987ApJ,yamada1998,machida2005},
\item (2) collapsing primordial gas in minihaloes with masses $\sim 10^{6}\ M_{\odot}$, which is substantially neutral during the process \citep{bromm1999,bromm2002}, and 
\item (3) initially ionized collapsing primordial gas in relic \HII\ regions, formed due to photo-ionization from first-generation massive stars \citep{yoshida2007}.
\end{itemize}
In each situation, the photo-destruction of molecular coolants is not taken into account, although this effect could be significant for $\mathrm{LiH}$ \citep{galli2013}, especially for cases (1) and (3), where Lyman-Werner (LW) and ionizing photons are likely important. Actually, including the interaction between baryons and the radiation background does not necessarily decrease the abundances of $\mathrm{H_{2}}$ and $\mathrm{HD}$, even though they can be destroyed by the photons. If the radiation spectrum contains enough energetic photons to ionize neutral hydrogen, the resulting free electrons catalyze the formation of $\mathrm{H_{2}}$ and $\mathrm{HD}$, which competes with the direct photo-destruction process (e.g. \citealt{johnson2007}). However, there is no such catalytic mechanism for $\mathrm{LiH}$, see reactions~(\ref{e9})-(\ref{e11}) below, so that the radiation field\footnote{To be more concrete, the radiation field can reduce the abundance of $\mathrm{LiH}$ when it contains a significant number of photons with energy above the dissociation energy of $\mathrm{LiH}$, given by $hcD_{e}(\mathrm{LiH})=2.515$~eV ($D_{e}(\mathrm{LiH})=20287.7~\mathrm{cm^{-1}}$, \citealt{stwalley1993}).} will definitely reduce the abundance of $\mathrm{LiH}$, and thus, reduce the contribution of $\mathrm{LiH}$ to cooling. Therefore, the results of our model can be regarded as the upper limit of the effect of $\mathrm{LiH}$ on the cooling of primordial gas.

The gas temperature, $T$, evolves with time according to (e.g. \citealt{machida2005}; \citealt[][GP98 henceforth]{galli1998}) 
\begin{align}
\frac{1}{\gamma-1}k_{\mathrm{B}}\frac{dT}{dt}=\frac{k_{B}T}{n}\frac{dn}{dt}+\frac{\Gamma-\Lambda}{n}\ ,\label{e1}
\end{align}
where $\gamma$ is the adiabatic index, equal to $5/3$ for the monatomic ideal gas assumed here, and $k_{\mathrm{B}}$ the Boltzmann constant. Further, $\Gamma-\Lambda$ denotes the net energy transfer from the radiation field to the gas, i.e. the heating minus cooling, and $n$ is the total number density of baryonic particles (i.e. ions, atoms, and molecules).
\ For simplicity, $\Gamma=0$ is assumed, as we neglect any photo-heating process in our modelling.

To derive the resulting thermal evolution, $T(t)$, we employ simple approximations for the dynamics of the system.
For case (1), near-isobaricity in the post-shock region is assumed, while for cases (2) and (3), we adopt a free-fall collapse model. The total gas number density thus evolves according to
\begin{align}
\frac{dn}{dt}=
\begin{cases}
&-\frac{n}{T}\frac{dT}{dt},\hspace{43pt} \text{isobaric}\\
&\frac{n}{t_{\text{ff}}},\hspace{60pt} \text{free-fall collapse.}
\end{cases}\label{e2}
\end{align}
Here, $t_{\text{ff}}=\left[3\pi/(32G\rho)\right]^{1/2}$ is the free-fall time, where $\rho=m n$ is the mass density, and $m=\mu m_{\mathrm{H}}$ the average mass of baryonic particles. The mean molecular weight is $\mu \simeq 4/(4-3Y)$, with $Y=0.24$ being the primordial He mass fraction.
To take into account the temperature floor, imposed by the CMB radiation, we adopt an effective cooling function for each molecular coolant as
\begin{align}
\Lambda=\Lambda(T)-\Lambda(T_{\mathrm{CMB}})\ ,\label{e3}
\end{align}
where $T_{\mathrm{CMB}}=2.73\times (1+z)\ \mathrm{K}$ is the CMB temperature.

\subsection{LiH cooling}
\label{s2.2}

The total cooling function for component X per unit volume (in units of $\mathrm{erg\ s^{-1}\ cm^{-3}}$) can be approximated (with a typical error of a factor of 2) as (e.g. \citealt{hollenbach1979,coppola2011})
\begin{align}
\Lambda_{\mathrm{X}}&=n(\mathrm{X})\cdot W_{\mathrm{X}}[T,n(\mathrm{collider})]\notag\\
&=n(\mathrm{X})\cdot\lambda_{\mathrm{X}}[T,n(\mathrm{collider})]\cdot n(\mathrm{collider})\ .\label{e4}
\end{align}
Here\footnote{Note that $W_{\mathrm{X}}$ (in $\mathrm{erg\ s^{-1}}$), the cooling rate per coolant particle, and $\lambda_{\mathrm{X}}$ (in $\mathrm{erg\ s^{-1}\ cm^{3}}$) are both sometimes called `cooling function', and the latter is occasionally denoted by the symbol $\Lambda_{\mathrm{X}}$ in the literature. Great care should be exercised in any comparison between papers.},
\begin{align}
&W_{\mathrm{X}}[T,n(\mathrm{collider})]=W_{\mathrm{X,LTE}}(T)\left[1+\frac{n_{\mathrm{X,cr}}(T)}{n(\mathrm{collider})}\right]^{-1}\ ,\label{e5}\\
&n_{\mathrm{X,cr}}(T)=\frac{W_{\mathrm{X,LTE}}(T)}{\lambda_{\mathrm{X},n\rightarrow 0}(T)}\ ,\label{e6}\\
&\lambda_{\mathrm{X},n\rightarrow 0}(T)=\lim_{n(\mathrm{collider})\rightarrow 0}\lambda_{\mathrm{X}}[T,n(\mathrm{collider})]\ ,\label{e7}
\end{align}
where $W_{\mathrm{X,LTE}}$ is the cooling rate per coolant particle under conditions of local thermodynamic equilibrium (LTE), $\lambda_{\mathrm{X},n\rightarrow 0}$ the low-density limit of the cooling rate, $n_{\mathrm{X,cr}}$ the critical number density for the transition to LTE, and $n(\mathrm{collider})$ the collider number density. Note that this formalism is the exact steady-state solution for a two-level system, when neglecting radiation fields, and can be used to express the cooling function in more realistic cases of enhanced complexity.

For $\mathrm{LiH}$, we only consider neutral hydrogen, $\mathrm{H^0}$, as the collider\footnote{According to \citep{shi2013}, in principle, collisions with other abundant species, such as $\mathrm{H_{2}}$, $\mathrm{He^0}$, $\mathrm{e^{-}}$ and $\mathrm{H^{+}}$, should be considered. Only using $\mathrm{H^0}$ introduces uncertainties for $n(\text{collider})<n_{\text{LiH,cr}}$, as vibrational excitation rate coefficients for $\mathrm{LiH-H^{0}}$ collisions are not available, and have to be estimated via mass-scaling from the $\mathrm{LiH-He^{0}}$ rate.}. After some straightforward rearranging of equations~(\ref{e4})-(\ref{e7}) above, we express the $\mathrm{LiH}$ cooling function as
\begin{align}
&\Lambda_{\mathrm{LiH}}=n(\mathrm{LiH})\cdot \frac{W_{\mathrm{LiH,LTE}}(T)\cdot n(\mathrm{H}^0)\lambda_{\mathrm{LiH},n\rightarrow 0}(T)}{W_{\mathrm{LiH,LTE}}(T)+n(\mathrm{H}^0)\lambda_{\mathrm{LiH},n\rightarrow 0}(T)}\ ,\label{e8}
\end{align}
where $n(\mathrm{LiH})$ is the number density of $\mathrm{LiH}$, and $n(\mathrm{H}^0)$ that of neutral hydrogen. We implement $W_{\mathrm{LiH,LTE}}$ from \citet{coppola2011}, and $\lambda_{\mathrm{LiH},n\rightarrow 0}$ from \citet{bougleux1997} and GP98. The cooling functions for the other molecular coolants are mainly the same with those employed in JB06. Specifically, the $\mathrm{H_{2}}$ cooling function is obtained from \citet{1987ApJ} for LTE, and from GP98 for the low-density case. We compare this fit of the $\mathrm{H_{2}}$ cooling function to the steady-state result from \citet{le1999}, based on results of quantum mechanical calculations. The deviation is within a factor of 2 for $T\lesssim 10^{3}\ \mathrm{K}$, and can reach up to a factor of 7 for $T>10^{3}\ \mathrm{K}$, which reflects the typical range of uncertainties (for details, see \citealt{glover2008}). 
For the $\mathrm{HD}$ cooling function, we adopt the $\lambda_{\mathrm{HD},n\rightarrow 0}$ in GP98 (accurate for $T\lesssim 10^{3}\ \mathrm{K}$), and the $W_{\mathrm{HD,LTE}}$ from \citet{coppola2011}, fitted for $10\ \mathrm{K}<T<3\times 10^{4}\ \mathrm{K}$. The resulting $W_{\mathrm{HD}}$, defined by formulas~(\ref{e5})-(\ref{e7}), is different from that adopted in JB06, based on \citet{flower2000}. The difference between these two versions of the $\mathrm{HD}$ cooling function is within a factor of 3 in the low temperature regime $T\lesssim 100\ \mathrm{K}$ that we are concerned with. We have verified that such a difference has minor impact on the thermal and chemical evolution, and will not affect the conclusions of this work.
Finally, the cooling functions for other processes, such as inverse Compton scattering, recombination, collisional ionization and excitation of helium and hydrogen species, are taken from \citet{machida2005}. 

At $T=100\ \mathrm{K}$, the critical densities of H$_2$ and HD are $n_{\mathrm{H_{2},cr}}\simeq 400\ \mathrm{cm^{-3}}$ and $n_{\mathrm{HD,cr}}\simeq 6000\ \mathrm{cm^{-3}}$, respectively, while that for LiH, $n_{\mathrm{LiH,cr}}\simeq 1.8\times 10^{10}\ \mathrm{cm^{-3}}$, is significantly larger, implying that the cooling efficiency (per molecule) of $\mathrm{LiH}$ is much higher than those of $\mathrm{H_{2}}$ and $\mathrm{HD}$ not only in the low-temperature regime, but also for high densities, as shown in the top panels of Figure~\ref{f1}. The high cooling efficiency of $\mathrm{LiH}$ can be attributed to its very large (ground-state) electric dipole moment of $d_{\mathrm{LiH}}=5.88$~D, compared with that of $\mathrm{HD}$, $d_{\mathrm{HD}}=8.3\times 10^{-4}$~D, where $1\ \mathrm{D}=10^{-18}\ \mathrm{esu\ cm}$ \citep{coppola2011}. The spontaneous-emission Einstein coefficient for the lowest-lying rotational transition of $\mathrm{LiH}$, i.e. the $v=0,\ J=1\rightarrow 0$ dipole transition within the $X^{1}\Sigma^{+}$ electronic state with $\Delta E_{10}/k_{B}=21.3\ \mathrm{K}$ \citep{bellini1995}, is $A_{10}(\mathrm{LiH})=1/\tau_{v}=1.18\times10^{-2}\ \mathrm{s^{-1}}$, where $\tau_{v}=84.49\ \mathrm{s}$ is the lifetime of the $v=0,\ J=1$ state \citep{juarros2006}. This is much larger than that of $\mathrm{HD}$, $A_{10}(\mathrm{HD})=5\times 10^{-8}\ \mathrm{s^{-1}}$ with $\Delta E_{10}/k_{B}=128\ \mathrm{K}$, as well as that of the H$_2$ quadrapole transition with $A_{20}(\mathrm{H_{2}})=3\times 10^{-11}\ \mathrm{s^{-1}}$ and $\Delta E_{20}/k_{B}=510\ \mathrm{K}$ \citep{nakamura2002}.

\subsection{Lithium chemistry}
\label{s2.3}

\begin{figure*}
\centering
\includegraphics[width=2.0\columnwidth]{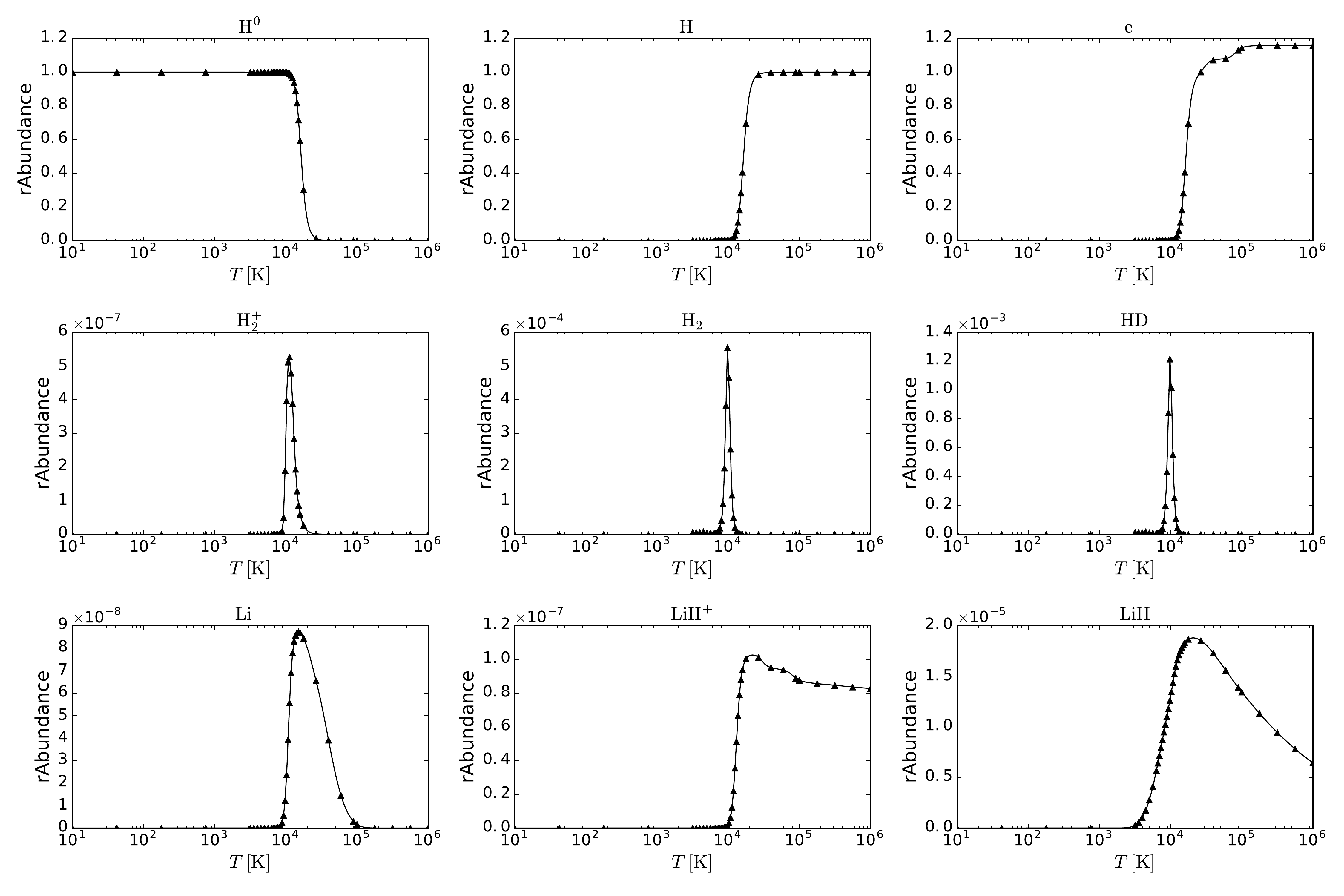}
\caption{Primordial chemistry. We show the equilibrium abundances, with respect to the corresponding total number density, of the relevant species in our chemical network. Solid lines denote analytical solutions, whereas triangles show the results of numerical calculations. It is evident that our numerical solver reproduces the analytical results extremely well.}
\label{f2}
\end{figure*}

We start with the chemical network in JB06, which includes hydrogen, helium and deuterium species \citep{haiman1996,galli1998,bromm2002,mackey2003} with a total deuterium abundance of $\mathrm{[D/H]=4\times 10^{-5}}$. In addition to those species, we consider the five lithium species $\mathrm{Li}^0$, $\mathrm{Li^{+}}$, $\mathrm{Li^{-}}$, $\mathrm{LiH^{+}}$ and $\mathrm{LiH}$, connected by the minimal model suggested by GP98 (see their fig. 6 and table 4). We neglect any reactions involving CMB photons, because they are much slower than other processes at the low CMB temperatures encountered here, $T_{\mathrm{CMB}}\simeq 15\ \mathrm{K}$ for $z=5$. In this simple chemical network of 10 reactions, $\mathrm{LiH}$ is formed by the following two channels
\begin{align}
&\mathrm{Li^{0}}\mathrm{+H^{0}\rightarrow LiH}+h\nu\ ,\label{e9}\\
&\mathrm{Li^{-}+H^{0}\rightarrow LiH+e^{-}}\ ,\label{e10}
\end{align}
where reaction~(\ref{e9}) is the main channel in the low-temperature regime that we are concerned with here. The rate coefficient for reaction~(\ref{e9}), i.e. the radiative association rate, is implemented in a state-resolved manner for the $^{2}S$ and $^{2}P$ states of atomic $\mathrm{Li^{0}}$, assuming that in LTE $[\mathrm{Li}^{0}(^{2}S)/\mathrm{Li^{0}}]=1-[\mathrm{Li}^{0}(^{2}P)/\mathrm{Li^{0}}]=1/[1+3\exp(-\delta_\epsilon/(k_{B}T))]$ , where $\delta_\epsilon=1.85\ $eV is the difference between the two energy levels \citep{galli2013}. It is important to point out that the rate coefficients for the $^{2}S$ and $^{2}P$ states in GP98 are based on quantum-mechanical calculations from \citet{dalgarno1996} and \citet{gianturco1996}, respectively, and that the resulting LTE rate is 2 to 3 orders of magnitude smaller than the semi-classical estimate used in the early study by \citet{1984ApJ} for $T\lesssim 10^{3}\ \mathrm{K}$.

The only channel for collisional $\mathrm{LiH}$ destruction is
\begin{align}
\mathrm{LiH+H^{0}\rightarrow Li^{0}+H_{2}}\ .\label{e11}
\end{align}
In GP98, the rate coefficient for this reaction adopts the estimate from \citet{stancil1996}, based on the exothermic energy between the $\mathrm{LiH+H^{0}}$ and $\mathrm{Li^{0}+H_{2}}$ states (2.258~eV, see \citealt{bovino2009}). The GP98 value is uncertain, and is smaller by up to one order of magnitude in the low-temperature regime that we are concerned with, $10\ \mathrm{K}\lesssim T\lesssim 100\ \mathrm{K}$, compared with more recent results from full quantum calculations (\citealt{bovino2009,2012JChPh}).
We have tested our chemical network by comparing the equilibrium abundances of the main species, calculated numerically by the above model, with those derived analytically, following the method in \citet{bromm2002}. As shown in Figure~\ref{f2}, the numerical results nicely reproduce the analytical solutions, and are consistent with those in JB06 (see their figure 1).

\section{Thermal and chemical evolution}
\label{s3}
\subsection{Maximal cooling}
\label{s3.1}
To ascertain whether it is possible for lithium hydride to play an important role in the cooling of primordial gas, we explore the case of maximal cooling in situations (1)-(3), by assuming that
\begin{itemize}
\item i) {\it all} lithium is locked up in the $\mathrm{LiH}$ molecule, and the chemical network for lithium species is turned off, 
\item ii) the total lithium abundance $\mathrm{[Li/H]}$ may be enhanced by one order of magnitude, with respect to its standard BBN value of $4.6\times 10^{-10}$ \citep{galli2013}, in non-standard models as described below.
\end{itemize}

The last assumption is based on non-standard BBN models, such as the decaying dark matter model of weakly interacting massive particles (WIMPs) (e.g. \citealt{jedamzik2008,jedamzik2009}). In this scenario, WIMPs are born from decaying massive particles during or after BBN, which also produce relativistic particles. These non-thermal, high-energy particles interact with the BBN plasma and change the light element abundances through disintegration, e.g. of $^{4}\mathrm{He}$. The $\mathrm{Li}$ abundance is affected via reactions with the resulting secondary particles, e.g. neutrons and deuterons, such as $\mathrm{n}+^{7}\mathrm{Be}\rightarrow \mathrm{p}+^{7}\mathrm{Li}$ and $\mathrm{d}+^{4}\mathrm{He}\rightarrow ^{6}\mathrm{Li}+h\nu$. Considering the decay of particle $\mathrm{X}$ with lifetime $\tau_{\mathrm{X}}$ and initial abundance $\zeta_{\mathrm{X}}=m_{\mathrm{X}}n_{\mathrm{X}}/n_{\gamma}$, $[\mathrm{Li/H}]\sim[\mathrm{^{7}Li/H}]$ can be enhanced by about one order of magnitude within a distinct region in $\tau_{\mathrm{X}}$-$\mathrm{\zeta_{\mathrm{X}}}$ space, i.e. $10^{3}<\tau_{\mathrm{X}}\ [\mathrm{s}]<10^{5}$ and $10^{-9}<\zeta_{\mathrm{X}}\ [\mathrm{Gev}]<10^{-7}$, as shown in figure 6 from \citet{fields2011}.

Previous work (e.g. JB06, \citealt{prieto2008}) has found that $\mathrm{HD}$ is the most important coolant at $T\lesssim 200\ \mathrm{K}$. To assess the impact of $\mathrm{LiH}$ cooling, it is thus appropriate to compare its efficiency with that of $\mathrm{HD}$ in the low-temperature regime. The bottom panels of Figure~\ref{f1} show how $W_{\mathrm{HD}}/W_{\mathrm{LiH}}$, the ratio of cooling rates per molecule, depends on the gas temperature and collider number density, indicating that the colder and denser the gas, the more efficient the cooling from $\mathrm{LiH}$ compared to $\mathrm{HD}$. {\it When does LiH cooling become competitive with HD?} In our calculation, closely matching the results of JB06 (see their figure 2), when the gas is initially fully ionized, the resulting HD abundance is $\mathrm{[HD/H]}\sim 10^{-6}$ at $T\lesssim 100\ \mathrm{K}$. If $\mathrm{[LiH/H]}$ assumes its extreme upper limit, equal to the primordial lithium abundance of $\sim 10^{-10}$ under assumption~i), to achieve $\Lambda_{\mathrm{HD}}/\Lambda_{\mathrm{LiH}}\lesssim 1$, the gas must evolve to a state of low temperature and high number density so that $W_{\mathrm{HD}}/W_{\mathrm{LiH}}\lesssim 10^{-4}$, while if we enhance the abundance of $\mathrm{LiH}$ by one magnitude, this condition becomes weaker,  $W_{\mathrm{HD}}/W_{\mathrm{LiH}}\lesssim 10^{-3}$.

\subsubsection{Shocked primordial gas}

\begin{figure*}
\centering
\includegraphics[width=2.0\columnwidth]{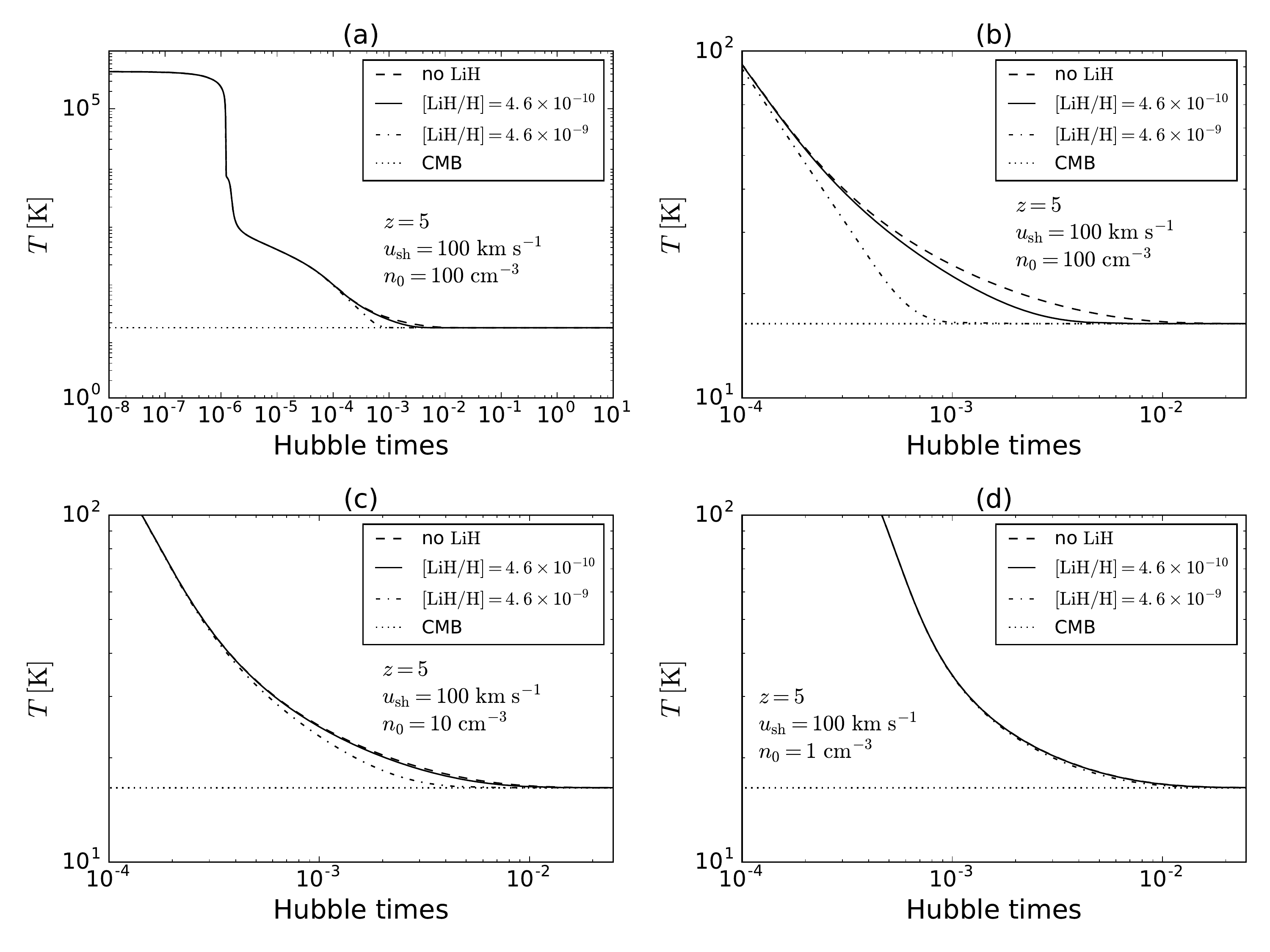}
\caption{Thermal evolution of shocked primordial gas under maximal $\mathrm{LiH}$ cooling. All cases have a shock velocity of $u_{\mathrm{sh}}=100\ \mathrm{km\ s^{-1}}$, occurring at redshift $z=5$. The CMB temperature is denoted by the dotted line. The initial number density, $n_{0}$, decreases from panel (a, b) $n_{0}=100\ \mathrm{cm^{-3}}$, to (c) $n_{0}=10\ \mathrm{cm^{-3}}$, and to (d) $n_{0}=1\ \mathrm{cm^{-3}}$. The last three panels only show the region where the temperature reaches the CMB floor. In each case, three values of the $\mathrm{LiH}$ abundance are considered: no $\mathrm{LiH}$ (dashed), $[\mathrm{LiH/H}]=4.6\times 10^{-10}$ (solid, the standard BBN value), and $[\mathrm{LiH/H}]=4.6\times 10^{-9}$ (dashed-dotted). It is evident that LiH cooling only marginally impacts the thermal evolution.}
\label{f3}
\end{figure*}

The primordial gas will be almost fully ionized in the radiative phase of the shocks with velocities $\gtrsim 100~\mathrm{km\ s^{-1}}$, originating from SN explosions of first-generation massive stars, or from cosmological structure formation \citep{kang1992,yamada1998,mackey2003,salvaterra2004}. The initial post-shock temperature of the gas is
\begin{align}
T_{0}=T_{\mathrm{ps}}=\frac{mu_{\mathrm{sh}}^{2}}{3k_{\mathrm{B}}}\sim 4.9\times 10^{5}\ \mathrm{K} \left(\frac{u_{\mathrm{sh}}}{100\ \mathrm{km\ s^{-1}}}\right)^{2}\ ,\label{e12}
\end{align}
where $u_{\mathrm{sh}}$ is the shock velocity. Here, a typical shock velocity $u_{\mathrm{sh}}=100\ \mathrm{km\ s^{-1}}$ is adopted, and cases with different values of the initial post-shock number density, $n_{0}$, as well as $\mathrm{LiH}$ abundance are studied. 

As shown in panel (a, b) of Figure~\ref{f3}, for $n_{0}=100\ \mathrm{cm^{-3}}$, the gas cools to the CMB floor in $\sim 10^{-2}t_{\rm H}$ without $\mathrm{LiH}$ cooling, where $t_{\rm H}=1.76\ \mathrm{Gyr}$ is the Hubble time at $z=5$ in the Planck cosmology\footnote{$\Omega_{m}=0.3089$, $H_{0}=67.74\ \mathrm{km\ s^{-1}}$ \citep{planck}.}. Cooling to the CMB floor is accelerated when LiH is present, to $\sim 4\times 10^{-3}t_{\rm H}$ and $\sim 10^{-3}t_{\rm H}$ for $\mathrm{[LiH/H]}=4.6\times 10^{-10}$, the standard BBN abundance, and $\mathrm{[LiH/H]}=4.6\times 10^{-9}$, respectively\footnote{Interestingly, even without $\mathrm{LiH}$ cooling, $\mathrm{HD}$ alone can cool the gas to the $z=5$ CMB floor of $T_{\mathrm{CMB}}<30\ \mathrm{K}$, even though the HD cooling efficiency drops rapidly for $T<30\ \mathrm{K}$.}. Given the same shock speed, for $n_{0}=10\ \mathrm{cm^{-3}}$, the effect of $\mathrm{LiH}$ is only marginally important for $\mathrm{[LiH/H]}=4.6\times 10^{-9}$, shifting the time taken for the temperature to reach the CMB floor from $\sim 10^{-2}t_{\rm H}$ to $\sim 5\times 10^{-3}t_{\rm H}$ (see panel (c) of Figure~\ref{f3}). While if $n_{0}=1\ \mathrm{cm^{-3}}$ (panel (d) of Figure~\ref{f3}), $\mathrm{LiH}$ makes no difference even if its abundance is enhanced by one order of magnitude, and this should also be the case for $n_{0}<1\ \mathrm{cm^{-3}}$. 
In general, Figure~\ref{f1} shows that $W_{\mathrm{HD}}/W_{\mathrm{LiH}}\sim 10^{-3}$ for $n(\mathrm{H^{0}})\sim 10^{6}\ \mathrm{cm^{-3}}$ at $T=40\ \mathrm{K}$, below which the $\mathrm{HD}$ cooling efficiency rapidly drops. Given the isobaric condition with post-shock pressure $P_{\rm ps}$, the initial values of density and temperature in the shocked gas need to fulfill the following criterion for $\mathrm{LiH}$ cooling to make any difference, assuming that the primordial abundance is boosted to $\mathrm{[LiH/H]\sim 4.6\times 10^{-9}}$
\begin{align}
\frac{P_\mathrm{ps}}{k_{\rm B}} &\simeq\left(\frac{n_{0}}{10^{6}\ \mathrm{cm^{-3}}}\right)\left(\frac{T_{0}}{40\ \mathrm{K}}\right)\notag\\
&\approx 1.25\left(\frac{n_{0}}{100\ \mathrm{cm^{-3}}}\right)\left(\frac{u_{\mathrm{sh}}}{100\ \mathrm{km\ s^{-1}}}\right)\gtrsim 1\ ,\label{cr}
\end{align}
which is marginally satisfied for the case of panel (a, b) in Figure~\ref{f3}. Here $P_{\mathrm{ps}}/k_{B}$ is normalized by the threshold value $4\times 10^{7}\ \mathrm{K\ cm^{-3}}$. 

\begin{figure*}
\centering
\includegraphics[width=2.0\columnwidth]{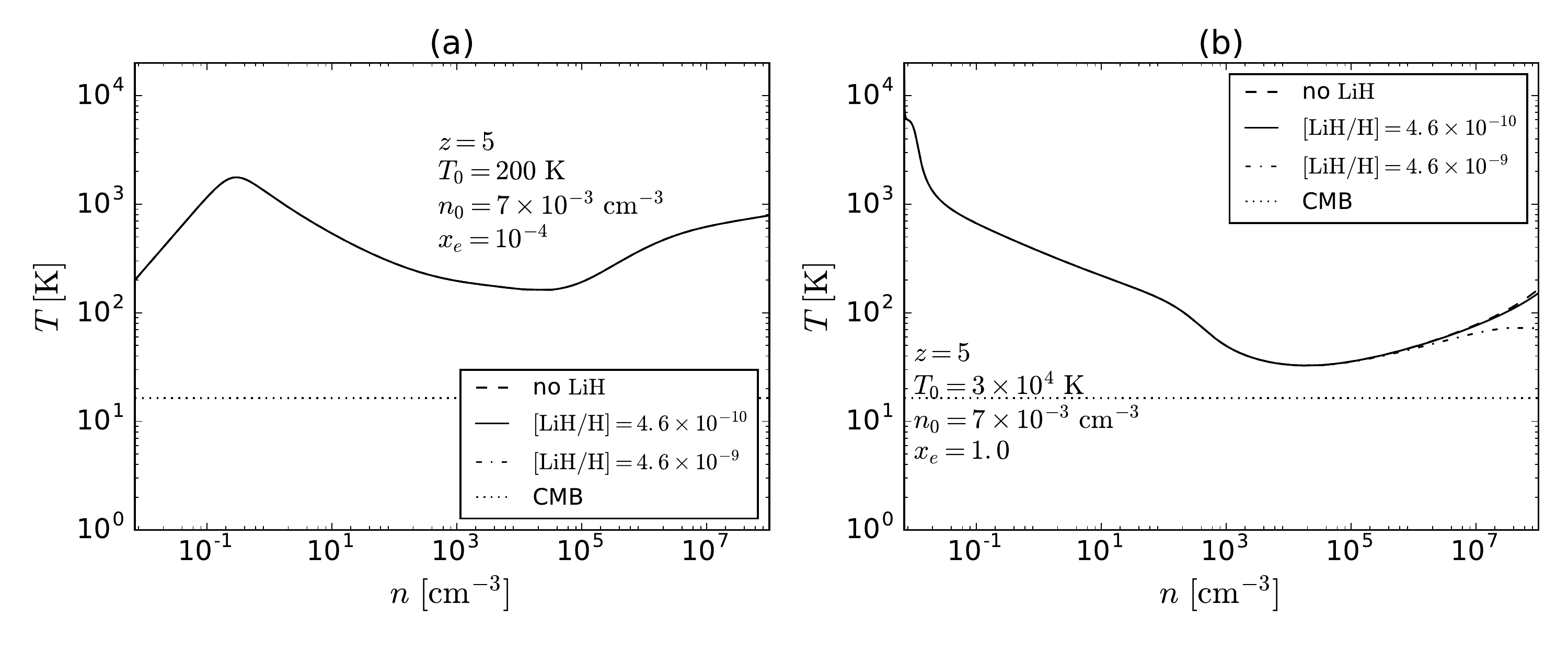}
\caption{Thermal evolution of freely-falling primordial gas, under maximal $\mathrm{LiH}$ cooling. {\it (a)}: Collapse into a minihalo, and {\it (b)} into a relic \HII\ region, at redshift $z=5$. For both cases, the initial number density of the gas is $n_{0}=7\times 10^{-3}\ \mathrm{cm^{-3}}$, and three values of the $\mathrm{LiH}$ abundance are considered: no $\mathrm{LiH}$ (dashed), $[\mathrm{LiH/H}]=4.6\times 10^{-10}$ (solid, the standard value), and $[\mathrm{LiH/H}]=4.6\times 10^{-9}$ (dashed-dotted). The initial gas temperature is $T_{0}=200\ \mathrm{K}$ for minihaloes with an initial ionization fraction of $x_{e}=10^{-4}$ ({\it panel a}), while that of the fully ionized gas in \HII\ regions ({\it panel b}) is $T_{0}=3\times 10^{4}\ \mathrm{K}$. In each case, the gas temperature does not reach the CMB floor, denoted by the dotted line.}
\label{f4}
\end{figure*}

\subsubsection{Unshocked primordial gas within minihaloes}
We next study the primordial gas, collapsing into a minihalo of mass $\sim 10^{6}\ M_{\odot}$. Before collapse, the initial temperature is $T_{0}=200\ \mathrm{K}$, the initial degree of ionization $x_{e}=10^{-4}$, and the initial number density $n_{0}$ is equal to the density of baryons in dark matter haloes at the point of virialization \citep{clarke2003}
\begin{align}
n_{0}\simeq 0.3\left(\frac{1+z}{21}\right)^{3}\ \mathrm{cm^{-3}}\ ,\label{e13}
\end{align}
which gives $n_{0}=7\times 10^{-3}\ \mathrm{cm^{-3}}$ for $z=5$. As shown in panel (a) of Figure~\ref{f4}, here the $\mathrm{LiH}$ cooling makes no difference at all, not even under the extreme maximal scenario.

\subsubsection{Ionized gas in relic \HII\ regions}
Considering the fully ionized primordial gas in a relic \HII\ region around a massive Pop~III star, the initial temperature is $T_{0}=3\times 10^{4}\ \mathrm{K}$ \citep{storey1995}, and the initial number density is the same as in the neutral minihalo case (see above), $n_{0}=7\times 10^{-3}\ \mathrm{cm^{-3}}$. As shown in panel (b) of Figure~\ref{f4}, the gas temperature drops by $\sim 100\ \mathrm{K}$ at $n\sim 10^{8}\ \mathrm{cm^{-3}}$ only when the primordial lithium abundance is boosted to $\mathrm{[LiH/H]\sim 4.6\times 10^{-9}}$. Furthermore, we find that this effect becomes even more significant at higher number density, $n\gtrsim 10^{8}\ \mathrm{cm^{-3}}$, at later stages of the collapse, consistent with the conclusion of \citet{1984ApJ}. Yet, this $\mathrm{LiH}$-enhanced cooling still cannot cool the gas to the CMB temperature, since heating by gravitational compression is strong. Besides, as the density increases, the gas starts to enter the molecular phase, so that our chemical network, without three-body reactions, is no longer complete \citep[e.g.][]{palla1983,turk2011}. In addition, opacity will begin to hinder the escape of the cooling radiation \citep[e.g.][]{greif2014,hirano2017}. Therefore, this regime is beyond the scope of the present study. \citet{2005PASJ} have investigated the evolution of spherical star-forming cores in this density regime, $n_{\mathrm{H}}\gtrsim 10^{9}\ \mathrm{cm^{-3}}$, taking into account three-body reactions for $\mathrm{LiH}$ formation\footnote{$\mathrm{Li^{0}+2H^{0}\rightarrow LiH+H^{0}}$ and $\mathrm{Li^{0}+H^{0}+H_{2}\rightarrow LiH+H_{2}}$.}, and the reverse dissociative reaction (see equation~(\ref{e11})), finding that $\mathrm{LiH}$ cooling is also insignificant there.

We also have explored higher values of $n_{0}$, up to $100\ \mathrm{cm^{-3}}$, in the minihalo and relic \HII\ region cases, only to find that the additional cooling by $\mathrm{LiH}$ still fails to drive the gas temperature closer to the CMB value.
The reason for this behaviour is as follows. For the primordial gas in minihaloes, without an initially high ionization fraction, the abundances of $\mathrm{H_{2}}$ and $\mathrm{HD}$ are always too low to cool the gas to $T\lesssim 100\ \mathrm{K}$, where $\mathrm{LiH}$ cooling would be significant. While in the case of relic \HII\ regions, the temperature always reaches its minimum $\sim 30\ \mathrm{K}$ at $n\sim n(\mathrm{H^{0}})\sim 10^{4}\ \mathrm{cm^{-3}}$, independent of $n_{0}$. From Figure~\ref{f1}, we know that under such conditions, $W_{\mathrm{HD}}/W_{\mathrm{LiH}}\sim 2\times 10^{-2}$, which does not satisfy the threshold criterion derived above.

\subsection{Cooling based on the chemical network}
\label{s3.2}
\begin{figure*}
\centering
\includegraphics[width=2.0\columnwidth]{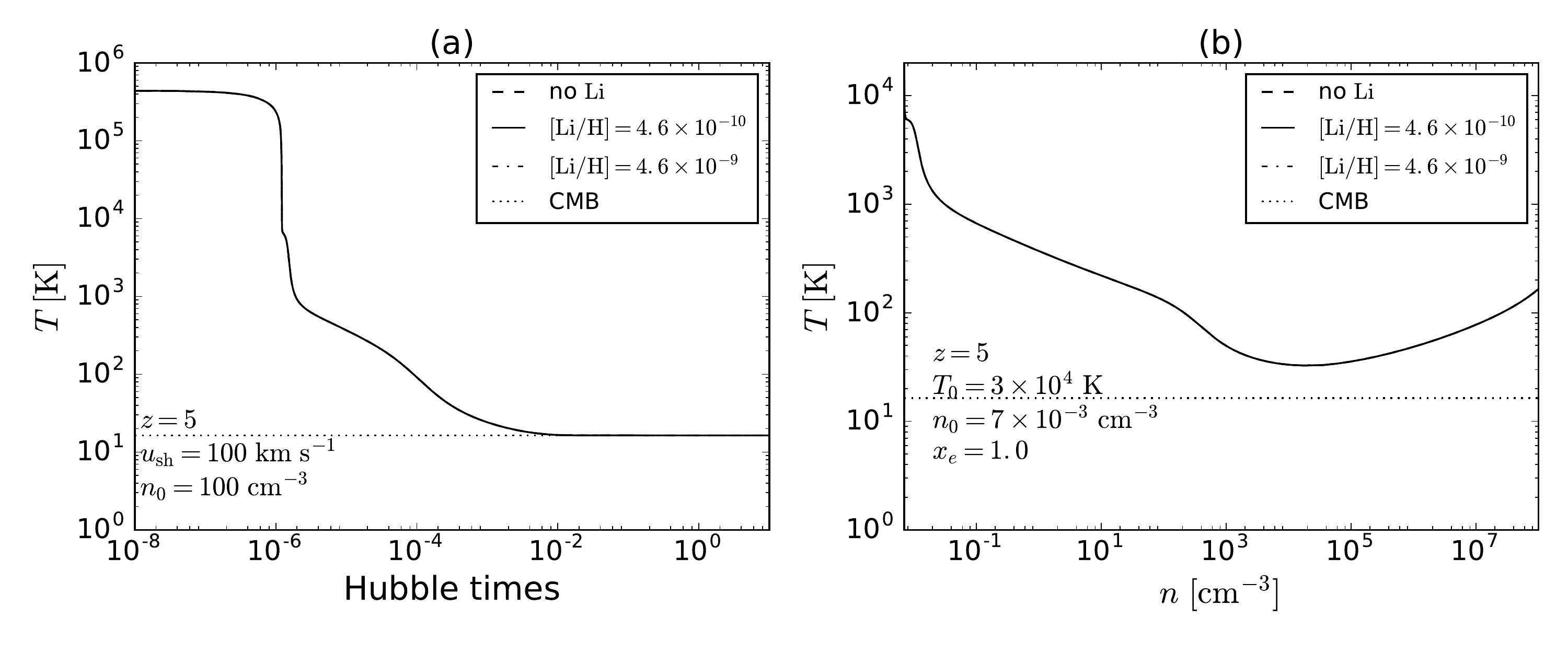}
\caption{Thermal evolution of primordial gas, evaluated with a chemical network for the lithium species. {\it (a)} Shocked isobaric gas, to be compared with the maximal cooling results in panel~(a) of Figure~\ref{f3}. {\it (b)} Collapse within a relic \HII\ region, to be compared with the maximal-cooling counterpart in panel (b) of Figure~\ref{f4}. As can be seen, under these more realistic conditions, $\mathrm{LiH}$ cooling has no effect on the overall evolution.}
\label{f5}
\end{figure*}

\begin{figure*}
\centering
\includegraphics[width=2.0\columnwidth]{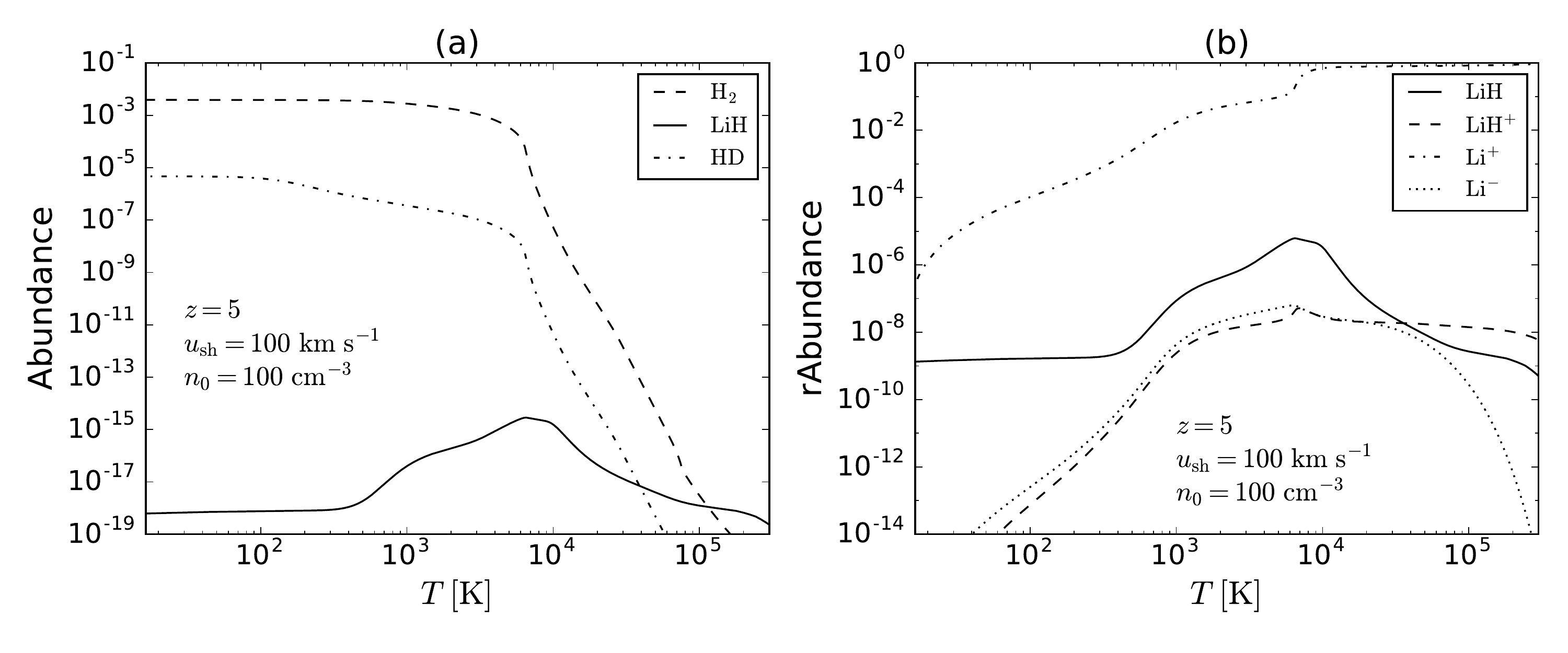}
\caption{Primordial abundances, in shocked gas. Here, the shock velocity is $u_{\mathrm{sh}}=100\ \mathrm{km\ s^{-1}}$, the initial post-shock number density $n_{0}=100\ \mathrm{cm^{-3}}$, and the redshift $z=5$. 
{\it (a)} Molecular coolant abundances, with respect to the total number of hydrogen nuclei: $\mathrm{H_{2}}$ (dashed line), $\mathrm{HD}$ (dashed-dotted) and $\mathrm{LiH}$ (solid). {\it (b)} Relative abundances, with respect to the total number of lithium nuclei, of key species: $\mathrm{LiH}$ (solid line), $\mathrm{LiH^{+}}$ (dashed), $\mathrm{Li^{+}}$ (dashed-dotted), and $\mathrm{Li^{-}}$ (dotted). The curve for $\mathrm{Li^{0}}$ is not plotted here, with a relative abundance close to one for $T\lesssim 10^{4}\ \mathrm{K}$. We note that the abundance of $\mathrm{LiH}$ remains low throughout under the more realistic scenario explored here, based on the full lithium chemical network.}
\label{f6}
\end{figure*}

\begin{figure}
\centering
\includegraphics[width=1.0\columnwidth]{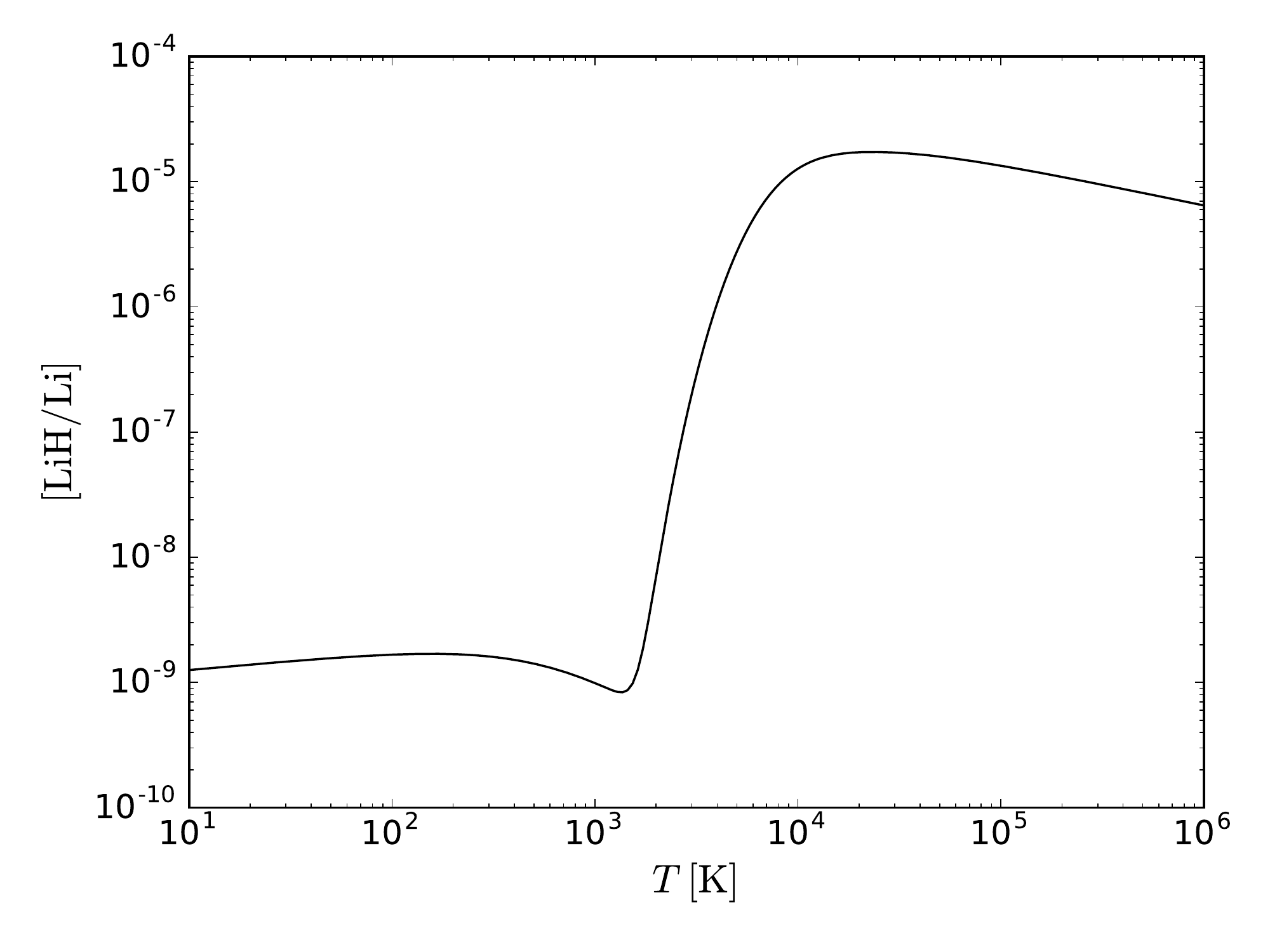}
\caption{Equilibrium $\mathrm{LiH}$ abundance, relative to the total number of Li nuclei, as a function of temperature $T$. We derive this ratio with the rates employed in our lithium network, specifically for the main $\mathrm{LiH}$ formation and destruction channels, i.e. reactions~(\ref{e9}) and (\ref{e11}). Again, it is evident that the abundance of $\mathrm{LiH}$ is extremely low in the relevant temperature range (see main text for details). 
}
\label{f7}
\end{figure}

The above results indicate that under maximal cooling, it is possible for $\mathrm{LiH}$ to make a marginal difference only in situations (1) and (3), where significant ionization activates the HD cooling channel. In this section, we revisit these two cases by considering the time evolution of the $\mathrm{LiH}$ abundance through our more realistic, minimal chemical network (see Section~\ref{s2.3}). Without loss of generality, we only consider the case with $n_{0}=100\ \mathrm{cm^{-3}}$ for situation (1).
The initial $\mathrm{LiH}$ abundance is set to zero, because its primordial value is negligibly small, $\sim 9\times 10^{-20}$ at $z=10$ \citep{galli2013}, and all lithium is initially in the form of $\mathrm{Li^{+}}$. 

Our results are displayed in Figure~\ref{f5}, to be compared with those in panel (a) of Figure~\ref{f3} and panel (b) of Figure~\ref{f4}. We find that now the effect of $\mathrm{LiH}$ cooling has entirely disappeared, which can be explained by the abundance evolution of the relevant lithium species, as shown in Figure~\ref{f6} for the standard $[\mathrm{Li/H}]=4.6\times 10^{-10}$. The resulting $\mathrm{LiH}$ abundance is extremely low, $[\mathrm{LiH/H}]\sim 10^{-19}$ at $T\lesssim 100\ \mathrm{K}$, and only reaches its maximum value of $\sim 10^{-15}$ at $T\sim 10^{4}\ \mathrm{K}$. In other words, as neutral lithium $\mathrm{Li^{0}}$ dominates for $T\lesssim 10^{4}\ \mathrm{K}$ and $[\mathrm{LiH/Li}]\sim 10^{-9}$ at $T\lesssim 100\ \mathrm{K}$, the abundance of $\mathrm{LiH}$ is reduced by nine orders of magnitudes compared with what we assume in our maximal cooling model, where {\it all} lithium is converted to $\mathrm{LiH}$. 
Such low $\mathrm{LiH}$ abundance is an intrinsic feature of our lithium chemical network.
Since the main channels of $\mathrm{LiH}$ formation and destruction, reactions~(\ref{e9}) and (\ref{e11}), share the same collider, $\mathrm{H^{0}}$, the equilibrium abundance ratio is  $\mathrm{[LiH/Li^{0}]}\simeq\mathrm{[LiH/Li]}=k_{\mathrm{cr}}/k_{\mathrm{des}}$, where $k_{\mathrm{cr}}$ is the rate coefficient of the formation reaction~(\ref{e9}), and $k_{\mathrm{des}}$ that of the destruction reaction~(\ref{e11}). As shown in Figure~\ref{f7}, the equilibrium $[\mathrm{LiH/Li}]$ ratio is in good agreement with the result in panel (b) of Figure~\ref{f6}, where we find $[\mathrm{LiH/Li}]\sim 10^{-9}$ for $T\lesssim 100\ \mathrm{K}$, and reaching a peak with $\sim 10^{-5}$ at $T\sim 10^{4}\ \mathrm{K}$. Thus, the $\mathrm{LiH}$ abundance, calculated by advancing the network of rate equations, is close to its equilibrium value. This is not the case for $\mathrm{H_{2}}$ and $\mathrm{HD}$, which are formed out of equilibrium in the low-temperature regime (see Figure~\ref{f2}). 
Moreover, taking into account the additional cooling from $\mathrm{LiH^{+}}$, whose cooling efficiency is around $10^{-2}$ that of $\mathrm{LiH}$ \citep{coppola2011}, will not change our results, because the abundance of $\mathrm{LiH^{+}}$ is even lower than that of $\mathrm{LiH}$.

\section{Summary and conclusions} 
\label{s4}
We investigate the effect of $\mathrm{LiH}$ on the cooling of primordial gas that remains uncontaminated to redshift $z=5$, in three characteristic environments: strong shocks, collapse into minihaloes, and collapse inside relic \HII\ regions. Under the extreme scenario of assuming that {\it all} lithium is converted to $\mathrm{LiH}$, and that the primordial lithium abundance is enhanced by one order of magnitude compared with the prediction of standard BBN, where $[\mathrm{Li/H}]=4.6\times 10^{-10}$, shocked gas can cool significantly faster to the CMB floor, given a high enough post shock pressure (see Sec.~\ref{s3.1}). 
While in the more realistic case where $\mathrm{LiH}$ is produced by the lithium chemical network, no effect is found, as the abundance of $\mathrm{LiH}$ is nine orders of magnitude smaller than the maximal limit. This is an intrinsic feature of the chemical network (see Sec.~\ref{s3.2}). Note that all assumptions and approximations made in this study, i.e. the relatively low redshift compared with the typical redshifts $z\gtrsim 10$ for early star formation, neglecting any photo-destruction effects, possibly underestimating the LiH dissociation rate, and the enhancement of the primordial lithium abundance, render our results a strong upper limit for the impact of $\mathrm{LiH}$ cooling.

We thus conclude that the effect of $\mathrm{LiH}$ on the cooling of primordial gas is always negligible for setting the initial conditions of fragmentation and star formation.
\ The baryonic aspects within the standard model of primordial star formation are thus complete, at least in terms of the relevant cooling processes. The remaining challenges lie in the arena of radiation-hydrodynamical feedback simulations \citep[e.g.][]{hosokawa2016,stacy2016}, and the even larger dependence on the unknown particle physics nature of the dark matter \citep[e.g.][]{hirano2018}. 

\subsection*{Acknowledgements}
This work was supported by NSF grant AST-1413501.

\bibliographystyle{mnras}
\bibliography{ref} 

\label{lastpage}
\end{document}